\documentstyle [preprint,aps,epsf,fixes]{revtex}

\def\eps{\epsilon}

\def\betaone{\beta^{(1)}}
\def\betatwo{\beta^{(2)}}

\def\asym{{\rm asym}}

\def\norm{{\cal N}}
\def\normu{\norm\!\mu^\eps}

\def\vconst{C}
\def\vconsta{\vconst_{11}}
\def\vconstb{\vconst_{10}}
\def\vconstc{\vconst_{\eps10}}

\def\m2{m^2}
\def\M2{M^2}
\def\Masym2{M_{\asym}^2}
\def\sqmu{\mu^2}
\def\pmu{\mu'}

\makeatletter\input eps1.aux
\begin {document}

\preprint {UW/PT-96-24}

\title  {$\eps$ expansion analysis of very weak first-order transitions
  in the cubic anisotropy model, Part II}

\author {Peter Arnold and Yan Zhang}

\address
    {%
    Department of Physics,
    University of Washington,
    Seattle, Washington 98195
    }%
\date {\today}
\maketitle

\begin {abstract}%
{%
  A companion article analyzed very weakly first-order phase transitions in
  the cubic anisotropy model using $\eps$ expansion techniques.  We extend
  that analysis to a calculation of the relative discontinuity of
  specific heat across the transition.
\ifpreprintsty
\thispagestyle {empty}
\newpage
\thispagestyle {empty}
\vbox to \vsize
    {%
    \vfill \baselineskip .28cm \par \font\tinyrm=cmr7 \tinyrm \noindent
    \narrower
    This report was prepared as an account of work sponsored by the
    United States Government.
    Neither the United States nor the United States Department of Energy,
    nor any of their employees, nor any of their contractors,
    subcontractors, or their employees, makes any warranty,
    express or implied, or assumes any legal liability or
    responsibility for the product or process disclosed,
    or represents that its use would not infringe privately-owned rights.
    By acceptance of this article, the publisher and/or recipient
    acknowledges the U.S.~Government's right to retain a non-exclusive,
    royalty-free license in and to any copyright covering this paper.%
    }%
\fi
}%
\end {abstract}

\section {Introduction}

   In this work, we use $\eps$ expansion methods to compute the universal
ratio $C_+/C_-$ of specific heats for arbitrarily weak first-order phase
transitions in the cubic anisotropy model.  $C_+$ and $C_-$ are the specific
heats of the disordered and ordered phases at the transition temperature.
This work is a direct follow up to
the computations of other universal ratios in ref.~\cite{susceptibility},
and we shall
eschew any discussion of motivation or review of method and notation;
instead, we jump directly into the calculation.  The reader should read
ref.~\cite{susceptibility} first.

   In the next section, we review the leading-order calculation of the
ratio, which was first performed by Rudnick in ref.~\cite{rudnick}.
Our result differs by a factor of 4.  In section \ref{sec:NLOcratio},
we proceed to next-to-leading order in $\eps$.
Our final results are displayed in section \ref{sec:cratiodiscussion}.


\section {Leading-order analysis of $C_+/C_-$}
\label{sec:LOcratio}

  Recall from the introduction of ref.~\cite{susceptibility} that,
in the three-dimensional
theory, the square $m^2$ of the scalar mass plays the roll of the reduced
temperature near the transition and the effective potential $V$ represents
the free energy of the system.
The specific heat
can be extracted from the effective potential as
\begin {equation}
   C \propto {d^2 V \over d (m^2)^2} \,.
\label{eq:Cpropto}
\end {equation}
The proportionality constant will not matter to our determination of the
ratio $C_+/C_-$.%
\footnote{
   In particular, the reduced temperature is proportional to $m^2(\mu_0)$
   for some fixed renormalization scale $\mu_0$, but we will instead apply
   (\ref{eq:Cpropto}) with $m^2(\mu)$ where $u\mu$ is roughly the order of
   the correlation length and varies as we approach the transition.
   [Specifically, we choose $\mu$ so that $u(\mu) = -v(\mu)$.]
   However, $m^2(\mu)$ is related to $m^2(\mu_0)$
   by multiplicative renormalization which, even though $\mu$ dependent,
   cancels in the ratio.
}

  Because of this relationship, it is useful to begin by summarizing the
leading-order form of the effective potential discussed in
ref.~\cite{susceptibility}.


\subsection{Summary of leading-order potential}

At one-loop order, the effective potential along an edge
$\vec\phi = (\phi,0,0,\cdots)$, evaluated at the tree-level instability 
line $u{=}-v$, is
[eq.~(\ref{V1 with m}) of ref.~\cite{susceptibility}]:
   \begin {eqnarray}
      \normu (V_0 + V_1)
      = \Lambda + 3 u^{-1} \m2 \M2
      &+& {1\over4} m^4 \left[
          \ln\left(\m2\over\sqmu\right) - {\vconstb\over\vconsta} \right]
   \nonumber\\
      &+& \vconsta(\m2+\M2)^2 \left[
          \ln\left(\m2+\M2\over\sqmu\right) - {\vconstb\over\vconsta} 
          \right] + O(\eps) \,,
   \label {eq:V01}
   \end {eqnarray}
where $\M2 \equiv {1\over6} \norm u \phi^2$, $\vconsta={1\over4}(n-1)$,
$\vconstb=-{3\over2}\vconsta$, and the normalization $\norm$ is
\begin {equation}
     \norm = (4\pi)^{d/2} \,\Gamma\!\left(\textstyle {d\over2} {-} 1\right)
\end {equation}
in $d=4{-}\epsilon$ spatial dimensions.

A first-order transition occurs as $m^2$ is varied.
At the transition, in the asymmetric phase, $m^2/M^2 \sim O(\eps)$.
In the asymmetric phase, the potential then reduces to
  \begin {equation}
     \normu (V_0 + V_1)
    = \Lambda + 3 u^{-1} \m2 \M2
      + M^4 \left[
              \vconsta \ln\left(\M2\over\sqmu\right) + \vconstb \right]
      + O(\eps^2 V)_\asym \,.
  \label {eq:V01asym}
  \end {equation}
The transition occurs when
  \begin{equation}
     m^2 = m_1^2 \left[1 + O(\eps) \right] \,,
  \end{equation}
and the asymmetric minimum is at
  \begin{equation}
     M^2  =  M_1^2 \left[1 + O(\eps) \right] \,,
  \label{eq:M2LO}
  \end{equation}
where
  \begin{eqnarray}
     m_1^2 & \equiv & {\vconsta u \sqmu\over3}
       \exp\left( -1 -{\vconstb\over\vconsta} \right)
     = {n{-}1 \over 12} \, u \, \mu^2 e^{1/2} \,,
  \label{eq:msq1}
  \\
     M_1^2 & \equiv & \sqmu
       \exp\left( -1 -{\vconstb\over\vconsta} \right)
     = \mu^2 e^{1/2} \,.
  \label{eq:Msq1}
  \end {eqnarray}
We will also need the variation of the asymmetric minimum as one varies
$m^2$ slightly away from the transition:
  \begin{equation}
     {d (M^2)\over d (\m2)} \biggr|_{m^2=m_1^2(1+O(\eps))} =
       - {3\over\vconsta u} ~+~ O(\eps^0) \,,
  \label{eq:dMdm1}
  \end{equation}

Finally, we will need the renormalization group flow of the various couplings,
previously presented in ref.~\cite{susceptibility}.  At leading order,
\begin{equation}
     \mu \partial_\mu \m2  = (\betaone_{\m2}(u,v) + O(u^2,v^2))\m2 \,,
\label{eq:m2rg1}
\end{equation}%
\begin{mathletters}%
\label{eq:couplingrg1}%
\begin {eqnarray}
     \mu \partial_\mu u & = & -\eps u + \betaone_u(u,v) + O(u^3,v^3)\,,
  \\
     \mu \partial_\mu v & = & -\eps v + \betaone_v(u,v) + O(u^3,v^3)\,,
\end{eqnarray}%
\end{mathletters}%
where
  \begin {eqnarray}
     \betaone_{\m2} &=& {n+2\over3} u + v \,,
  \label{eq:betaonem2}
  \\
     \betaone_u &=& u\left({n+8\over3} u + 2 v\right) \,,
  \\
     \betaone_v &=& v(4u + 3v) \,.
  \label{eq:betaone}
  \end {eqnarray} 
It is also convenient to introduce $f \equiv u/v$ and
\begin {equation}
  \mu \partial_\mu f = \betaone_f(u,f) + O(u^2,v^2)\,,
\label{eq:betaf1}
\end {equation}
\begin {equation}
      \betaone_f = u\left({n-4\over3}f - 1\right) \,.
      \label{eq:betaonef}
\end {equation}


\subsection{Specific heats}
 
  The most non-trivial task in computing the specific heat ratio at the
transition will be to handle the renormalization group flow of the
constant term $\Lambda$ in the potential.  Before investigating this,
we first write down the expression for $C_+$ and $C_-$ given the
potential.  From (\ref{eq:V01}),
  \begin{eqnarray}
     C_+ &\propto& \left(d \over d (\m2) \right)^2
                     \normu V(\M2{=}0) |_{\m2=m_1^2(1+O(\eps))} 
  \nonumber\\
         & = & \left(d \over d (\m2) \right)^2 \Lambda
              + {n\over2} 
                \left[\ln \left(m_1^2\over\sqmu\right) 
                \right] + O(\eps)
  \nonumber\\
         & = & \left(d \over d (\m2) \right)^2 \Lambda
              + 2\left({1\over4} + \vconsta\right) 
                \left[\ln \left(2\vconsta u\right)
                      + {1\over2} \right] + O(\eps) \,, 
  \label{eq:cplus1}
  \end{eqnarray}
where the second derivative of $\Lambda$ is understood to be evaluated
at the transition $m^2 = m_1^2(1+O(\eps))$.
Using (\ref{eq:V01asym}) for the asymmetric phase,
and remembering that $\partial_M V(M) = 0$ at the asymmetric minimum,
  \begin{eqnarray}
     C_- &\propto& \left(d \over d (\m2) \right)^2 
                    \normu V(M)|_{\m2=m_1^2(1+O(\eps))}
  \nonumber\\
         & = & \left(d \over d (\m2) \right)^2\Lambda
             + {3\over u} {d (M^2) \over d(\m2)}
             \biggr|_{\m2=m_1^2} + O(1/\eps)
  \nonumber\\
         & = & \left(d \over d (\m2) \right)^2\Lambda
              - {9\over u^2} {1 \over \vconsta} + O(1/\eps) \,,
  \label{eq:cminus1}
  \end{eqnarray}
where $M$ has been evaluated at the transition using
(\ref{eq:M2LO}) and (\ref{eq:dMdm1}).
$C_+$ will be shown to be $O(1/\eps)$ while $C_-$ is
$O(1/\eps^2)$.


\subsection{The running of $\Lambda$}

  Now we turn to the contribution to the specific heat from the
term $\Lambda$.  The renormalization group equation for $\Lambda$ 
at one-loop order is
  \begin {equation}
     \mu \partial_\mu\Lambda = \eps \Lambda + {n\over2} m^4 
                               (1 + O(u,v)) \,.
  \label{eq:lambdarg1}
  \end {equation}
The solution, in terms of $m^2(\mu)$, is
  \begin{equation}
     \Lambda(\mu) 
    = \left(\mu\over\mu_0\right)^{\eps} \Lambda(\mu_0)
       + {n\over2} \int^\mu_{\mu_0} {d \pmu\over\pmu}
         \left(\mu\over\pmu\right)^{\eps} [\m2(\pmu)]^2 (1 + O(u,v)) \,, 
  \label{eq:lambdrun1}
  \end{equation}
where $\mu_0$ is some initial scale.  The running (\ref{eq:m2rg1}) of $\m2$
yields
  \begin{equation}
     \m2(\mu) = E^{(1)}(\mu,\mu_0) \, \m2(\mu_0) \, (1 + O(u,v)) \,,
  \end{equation}
where
  \begin{equation}
     E^{(1)}(\mu,\mu_0) = \exp\left(\int^\mu_{\mu_0} {d\pmu\over\pmu} 
                              \betaone_{\m2}\right)
  \end{equation}
and the integral is to be understood as evaluated along the
leading-order solution for the coupling constant trajectory.
The contribution of $\Lambda$ to the specific heat is then
  \begin{equation}
     \left(d \over d (\m2) \right)^2 \Lambda 
    = n \int^\mu_{\mu_0} {d\pmu\over\pmu} 
      \left(\mu\over\pmu\right)^{\eps} 
      \left[E^{(1)}(\pmu,\mu)\right]^2 (1 + O(u,v)) \,.
  \label{eq:clambda1}
  \end{equation} 
$E^{(1)}$ can be easily evaluated by changing variables from $\mu'$ to
$f$ using (\ref{eq:betaf1}) and noting that $\betaone_{m^2}/\betaone_f$
(\ref{eq:betaonem2},\ref{eq:betaonef}) depends only on $f$:
  \begin{equation}
     E^{(1)}(\mu,\mu_0) = \exp\left(\int^{f}_{f_0}
                              {df'\over \betaone_f} \betaone_{\m2} \right)
    = {f_0\over f}
      \left( f_0+\lambda \over f+\lambda \right)^{(n\lambda-1)/2}
    \,,
  \label{eq:E1}
  \end{equation}
where
  \begin {equation}
     \lambda \equiv {3\over 4-n} \,.
  \label{eq:lambdef}
  \end {equation}

The remaining integral in (\ref{eq:clambda1}) can be performed similarly
if we make use of the following relation
((\ref{eq:uv solns}a) of ref.~\cite{susceptibility})
for the solution to the leading-order RG flow equations
(\ref{eq:couplingrg1}):
   \begin {equation}
   \left(\mu\over\mu_0\right)^\eps =
                     {u_0 \over u} \left(f_0\over f\right)^2
                     \left(f_0 + \lambda \over f+\lambda\right)^{n\lambda} .
   \end {equation}
This gives
  \begin{eqnarray}
     \left(d \over d (\m2) \right)^2 \Lambda 
    &=& -{n\lambda\over u(f+\lambda)} \int^f_{f_0} df' 
      \left(f'\over f\right)^2
             \left(f'+\lambda\over f+\lambda\right)^{n\lambda-1}
      \left[E^{(1)}(f',f)\right]^2 (1 + O(u,v))
  \nonumber\\
    &=& -{n\lambda\over u(f+\lambda)} \int^f_{f_0} df' (1 + O(u,v)) 
    \,.
  \end{eqnarray} 
So the result is
  \begin{equation}
     \left(d \over d (\m2) \right)^2 \Lambda 
    = -{n \lambda \over u (\mu)}
       {[f(\mu) - f(\mu_0)] \over (f(\mu) + \lambda)}\, [1 + O(\eps)] \,.
  \label{eq:clambdaf1}
  \end{equation}

  The only other elements we need are the values of $f(\mu_0)$, $f(\mu)$,
and $u(\mu)$ for the desired trajectory.
As discussed in ref.~\cite{susceptibility},
we can obtain the universal ratios of interest
by studying the trajectory that flows away from the cubic fixed point
at $f(\mu_0) = -\lambda$ to the line of classical instability
$u{=}-v$ at $f(\mu)=-1$ and $u=u_\star$, where, at leading order,
  \begin{equation}
     u_\ast = {3 (n^2+5n+3) \over n (n+2) (n+8)} \eps \,.
  \label{eq:ustr1}
  \end {equation}
For this trajectory,
  \begin{equation}
     \left(d \over d (\m2) \right)^2 \Lambda 
    = - {n\lambda\over u_*} + O(\eps^0) \,.
   \label{eq:dlam1}
  \end{equation}
This is $O(1/\eps)$ and dominates $C_+$ (\ref{eq:cplus1}), but it does
not contribute at leading order to $C_-$ (\ref{eq:cminus1}).
Putting it all together,
  \begin{eqnarray}
     C_+ \over C_- 
   &=& {n \lambda \vconsta \over 9} u_\ast ~+~ O(\eps^2) 
  \nonumber\\
   &=& {(n-1)(n^2+5n+3)\over4(n+2)(n+8)(4-n)} \eps 
       ~+~ O(\eps^2)   
  \nonumber\\
   &=& {17\over320}\eps ~+~ O(\eps^2) \qquad {\rm for}\;\; n=2 \,.
  \label{eq:cratio1}
  \end{eqnarray}
This result is $4$ times smaller than the result originally quoted by
Rudnick \cite{rudnick}.


\section {Next-to-leading-order analysis of $C_+/C_-$}
\label{sec:NLOcratio}

Because the contribution of $\Lambda$ dominated $C_+$, our formula
(\ref{eq:cplus1}) for $C_+$ is adequate at next-to-leading order
provided we extend our analysis of $\Lambda$ to next-to-leading order.
Before doing so, let us first consider $C_-$.


\subsection{The asymmetric phase: $C_-$}

  The two-loop effective potential near the asymmetric phase is
given in ref.~\cite{susceptibility} (see eqs.~(\ref{eq:v0 plus v1}), 
(\ref{eq:delta V1}), and (\ref{eq:V2 form})):
  \begin{eqnarray}
     \normu (V_0 + V_1 + V_2)
    = \Lambda & + & 3 u^{-1} \m2 \M2 + 
      M^4 \left[\vconsta \ln\left(\M2\over\sqmu\right) + \vconstb \right]
  \nonumber\\
      & + &\normu \delta V_1 + \normu V_2
            + O(\eps^2 V)_\asym \,,
  \label{eq:V012asym}
  \end{eqnarray}
where
  \begin{eqnarray}
     \normu \, \delta V_1
   &=& 2 \vconsta \m2 \M2 \left[\ln\left(\M2\over\sqmu\right) - 1 \right]
  \nonumber\\ 
   && \qquad 
      + \eps M^4 \left[-{1\over4}\vconsta \ln^2 \left(\M2\over\sqmu\right)
                       -{1\over2}\vconstb \ln \left(\M2\over\sqmu\right)
                       + \vconstc \right] \,,
  \label {eq:deltaV1asym}
  \\
     \normu V_2
    &=& u M^4 \left[\vconst_{22}\ln^2\left(\M2\over\sqmu\right)
                  + \vconst_{21}\ln\left(\M2\over\sqmu\right)
                  + \vconst_{20} \right] \,,
  \label{eq:V2asym}
  \end{eqnarray}
and
   \begin {equation}
      \vconst_{22} =  {n+2\over6} \vconsta \,,
      \qquad
      \vconst_{21} = -{(n+6)\over3} \vconsta \,.
   \label{eq:vconst2}
   \end {equation}
As discussed in ref.~\cite{susceptibility}, we shall not need to know
$\vconstc$ and $\vconst_{20}$.
The transition takes place at
  \begin{equation}
     \m2 = m_2^2 \left[1 + O(\eps^2) \right] \,,
  \end{equation}
with the asymmetric minimum at
  \begin{equation}
     M^2 = M_2^2 \left[1 + O(\eps^2) \right] \,,
  \end{equation}
where
  \begin {eqnarray}
     m_2^2 &=& m_1^2 \left[
       1 + \left(-{5\over16} - {\vconstc\over\vconsta}\right)\eps
         + \left({(n-4)\over24} - {1\over2} {\vconst_{21}\over\vconsta}
              - {\vconst_{20}\over\vconsta}\right)u \right] \,,
  \\
     M_2^2 &=& M_1^2 \left[
       1 + \left(-{13\over16} - {\vconstc\over\vconsta}\right)\eps
         + \left(-{(3n+2)\over8} - {3\over2} {\vconst_{21}\over\vconsta}
              - {\vconst_{20}\over\vconsta}\right)u \right] \,.
  \label{eq:msq2Msq2}
  \end {eqnarray}
At the transition,
  \begin{equation}
     {d (M^2)\over d (\m2)} \biggr|_{m^2=m_2^2(1+O(\eps^2))} =
       - {3\over\vconsta u}
                 \left[1+(\vconsta - 3 {\vconst_{22} \over \vconsta}
                        - {\vconst_{21} \over \vconsta})
                       + O(\eps^2) \right]
      \,.
  \label{eq:dMdm2}
  \end{equation}
The result for $C_-$ is
  \begin{eqnarray}
     C_- &\propto& \left(d \over d (\m2) \right)^2 
                   \normu V(M)
                    |_{\m2=m_2^2(1+O(\eps^2))}
  \nonumber\\
         & = &\left(d \over d (\m2) \right)^2 
              \Lambda
              + {d(M^2) \over d (\m2)} \left[ {3\over u} + 
              + 2 \vconsta \ln \left(M^2\over\sqmu \right) \right]
                \Biggr|_{\m2=m_2^2} + O(\eps^0) 
  \nonumber\\
         & = & \left(d \over d (\m2) \right)^2 
               \Lambda
              - {9\over u^2} {1 \over \vconsta}
                 \left[1+({4\over3}\vconsta - 3 {\vconst_{22} \over \vconsta}
                        - {\vconst_{21} \over \vconsta}) u \right]
              + O(\eps^0) \,.
  \label{eq:cminus2}
  \end{eqnarray}
Since the contribution of $\Lambda$ is sub-leading, the leading-order
result (\ref{eq:dlam1}) for the contribution is adequate here.


\subsection{The NLO running of $\Lambda$}

  Now we are left with calculating $d^2\Lambda/d(m^2)^2$ to next-to-leading
order.  The two-loop renormalization group equation is
  \begin {equation}
     \mu \partial_\mu \Lambda 
    = \eps \Lambda + {n\over2} (\m2)^2(1 + O(u^2,v^2)) \,,
  \label{eq:lamdrg2}
  \end {equation}
which has the same form as the one-loop equation.  It's solution then
also has the same form,
  \begin{equation}
     \left(d \over d (\m2) \right)^2 \Lambda 
    = n \int^\mu_{\mu_0} {d \pmu\over\pmu} \left(\mu\over\pmu\right)^{\eps}
      [E(\pmu,\mu)]^2(1 + O(u^2,v^2)) \,,
  \label{eq:clambda2}
  \end{equation}
where
  \begin{equation}
     E(\mu,\mu_0) = \exp\left(\int^\mu_{\mu_0} {d\pmu\over\pmu} 
                             \beta_{\m2}\right) \,,
  \label{eq:m2run2}
  \end{equation}
but now both the $\beta$-functions and renormalization group trajectories
should be evaluated at two loops.
The two-loop $\beta$-functions are given in 
Sec.~\ref{sec:twoloopRG} of ref.~\cite{susceptibility}:
  \begin{equation}
     \beta = \betaone + \betatwo + O(u^4,v^4) \,,
  \label{eq:beta12}
  \end{equation}
where
  \begin{eqnarray}
     \betatwo_{m^2} &=& - {5\over6} \left[
           {(n+2)\over3} u^2 + 2 u v + v^2 \right] \,,
  \\
     \betatwo_u &=&
         - {(3n+14)\over3} u^2 - {22\over3} u^2 v - {5\over3} u v^2 \,,
  \\
     \betatwo_v &=&
         - {(5n+82)\over9} u^2 v - {46\over3} u v^2 - {17\over3} v^3 \,.
  \end{eqnarray}

To evaluate the integrals, we again change variables from $\mu$ to $f$,
and we shall treat the two-loop effects on $\beta$-functions and
trajectories perturbatively.  Following sec.~\ref{sec:ustar2} 
of ref.~\cite{susceptibility}, it is helpful to make the $\eps$
dependence explicit by rewriting $(u,v) = \eps(\bar u,\bar v)$,
and the expansion of the trajectory gives
  \begin {equation}
     \bar u(f) = \bar u^{[1]}(f) + \eps\,\delta(f) + O(\eps^2) \,,
  \end {equation}
where $\bar{u}^{[1]}(f)=fR(f,c)$ is the one-loop result described
in sec.~\ref{sec:LOflow} of ref.~\cite{susceptibility}.
The solution for $\delta(f)$ is given by eq.~(\ref{eq:deltaf}) of
ref.~\cite{susceptibility}.

To change variables from $\mu$ to $f$, we use
\begin {eqnarray}
   {d\mu\over\mu} &=& {d f \over \beta_f}
    = \left[\betaone_f + \eps \betatwo_f + O(\eps^2)\right]^{-1}
           \biggr|_{\bar{u}(f),f}
\nonumber\\ &&
    = {d f\over\betaone_f} \left[1
           -{\eps\over\betaone_f}
            \left(\delta(f)\partial_{\bar u}\betaone_f + \betatwo_f\right)
           + O(\eps^2)\right]
           \biggr|_{\bar{u}^{[1]}(f),f} \,.
\label{eq:fexpansion}
\end {eqnarray}
The subscript ``$\bar{u}^{[1]}(f),f$'' at the end of this
equation means that the $\beta$-functions in the expression are to be
evaluated with $u\to\bar u^{[1]}(f)$ and $v \to\bar u^{[1]}(f)/f$.
Expanding the definition (\ref{eq:m2run2}) in $\eps$ then gives
\begin {equation}
     E(f, f_0)
    = E^{(1)}(f, f_0)\, [1 + \eps \delta E(f, f_0) + O(\eps^2)] \,,
  \label{eq:E2}
  \end{equation}
where $E^{(1)}$ is the leading-order form of (\ref{eq:E1}) and
  \begin{eqnarray}
     \eps \delta E(f, f_0)
    =\int_{f_0}^f d f' 
     &&\left\{ {1\over\betaone_f} 
              \left[\delta(f') \partial_{\bar{u}} \betaone_{\m2} 
                    + \betatwo_{\m2} \right] \right.
  \nonumber\\ && \qquad
       \left. - {\betaone_{\m2}\over \left(\betaone_f\right)^2}
              \left[\delta(f')\partial_{\bar{u}}\betaone_f 
                    + \betatwo_f \right]
       \right \} \Biggr|_{\bar{u}^{[1]}(f'),f'} \,.
  \label{eq:deltaE}
  \end{eqnarray}

To do the final integral of (\ref{eq:clambda2}), we need an expansion
of $(\mu/\mu')^\eps$.  This can be obtained by writing
\begin {equation}
   \left(\mu\over\mu'\right)^\eps
     = \exp\left(\eps\int_{\mu'}^\mu {d\mu''\over\mu''}\right) \,,
\end {equation}
and converting to $f$ with the expansion (\ref{eq:fexpansion}).
Putting all the expansions together yields
  \begin{eqnarray}
     \left(d \over d (\m2) \right)^2 \Lambda 
    &=& -{n\lambda\over u^{[1]}(f(\mu)) \, (f(\mu)+\lambda)}
  \nonumber\\ && \qquad\times
      \int_{f(\mu_0)}^{f(\mu)} d f'
       \left \{ 1 + \eps \left[ 2 \delta E(f',f(\mu))
                  + X_1(f') + X_2(f') \right] 
                + O(\eps^2) \right \} \,,
  \label{eq:clambdaf2}
  \end{eqnarray}
where
  \begin{equation}
     X_1(f') = - {1\over\betaone_f}
                   \left(\delta(f')\partial_{\bar{u}}\betaone_f 
                         + \betatwo_f\right)
            \Biggr|_{\bar{u}^{[1]}(f\prime),f\prime} \,,
  \label{eq:X1}
  \end{equation}
  \begin{equation}
     X_2(f') = - \int_{f'}^{f(\mu)}
           {d f'' \over\left(\betaone_f\right)^2}
                 \left(\delta(f'')\partial_{\bar{u}}\betaone_f 
                       + \betatwo_f \right)
            \Biggr|_{\bar{u}^{[1]}(f''),f''} \,.
  \label{eq:X2}
  \end{equation}
This integration does not seem to have a simple form for general $n$.
For $n=2$, we are able to obtain a simple result for the trajectory
flowing away from the cubic fixed point to the classical instability
line $u=-v$:
  \begin{equation}
     \left(d \over d (\m2) \right)^2 \Lambda
    = -{3\over u_\ast}\left\{1+\eps\left[{49\over40} 
                                              -{3\over5}\ln{3\over2}
                                       \right] + O(\eps^2) \right \}\,,
  \label{eq:clambdanum2}
  \end{equation}

Putting everything together, and using the next-to-leading order $n{=}2$
result
  \begin {equation}
     u_\ast = {51\over80}\eps + \left({243\over80}\ln{3\over2}
                                      - {171\over200} \right) \eps^2
                              + O(\eps^3)
  \label{eq:ustr2}
  \end {equation}
for $u_\ast$ from eq.~(\ref{eq:NLO ustar}) of ref.~\cite{susceptibility},
our final result is then
  \begin{equation}
     {C_+ \over C_-} 
    = {17 \over 320} \eps 
      \left\{1 - {17\over80} \eps \ln \eps 
              + \eps \left[{17\over80} \ln {320\over17} 
                           + {354\over85} \ln {3\over2} 
                           - {4967\over5440} \right]
              +O(\eps^2) \right\}
   \label{eq:cratio2}
   \end{equation}                   
for $n{=}2$.                         


\section {Discussion}
\label{sec:cratiodiscussion}

  Evaluated numerically, the final result (\ref{eq:cratio2}) for the
ratio is
  \begin{equation}
     {C_+ \over C_-}
    = 0.0531 \eps \left[1 + \eps(- 0.2125 \ln \eps + 1.3993)
                          + O(\eps^2) \right] \,.
   \end{equation}
This ratio is compared against Monte Carlo simulations \cite{numerical} 
in ref.~\cite{summary}.  The 140\% correction at next-to-leading order
for $\eps{=}1$ suggests that the $\eps$ expansion will be at best
marginally successful for this quantity.

\bigskip

This work was supported by the U.S. Department of Energy,
grants DE-FG06-91ER40614 and DE-FG03-96ER40956.
We thank Larry Yaffe for useful discussions.


\end {document}